\documentclass{iopjournal}

\usepackage[english]{babel}
\usepackage[utf8]{inputenc}
\usepackage{amsmath}
\usepackage{csquotes}
\usepackage{tabularx}
\usepackage{booktabs}
\usepackage{amsfonts}
\usepackage{mathtools}
\usepackage{multirow}
\usepackage{subcaption}
\usepackage[nolist,nohyperlinks]{acronym}
\usepackage{cleveref}
\usepackage{bm}
\usepackage{siunitx}
\usepackage[backend=bibtex,style=authoryear,maxbibnames=10]{biblatex}
\usepackage{tikz}
\usetikzlibrary{positioning}

\usepackage{tikz}
\usepackage{pgfplots}

\pgfplotsset{compat=1.18}

\usepgfplotslibrary{units}
\usepgfplotslibrary{colorbrewer}    
\usetikzlibrary{calc,positioning}
\usetikzlibrary{fit, backgrounds}   
\pgfplotsset{
    unit markings=slash space,
}

\tikzset{
	relative to node/.style={
        shift={(#1.center)},
        x={(#1.east)},
        y={(#1.north)}, 
    }
}

\pgfdeclarelayer{background}
\pgfdeclarelayer{foreground}
\pgfsetlayers{background,main,foreground}

\usepackage{color}
\definecolor{ibilight}{RGB}{193,216,237}
\definecolor{ibidark}{RGB}{0,73,146}	
\definecolor{uke2}{RGB}{170,156,143} 	
\definecolor{uke3}{RGB}{87,87,86}		
\definecolor{ukesec1}{RGB}{255,223,0}	
\definecolor{ukesec2}{RGB}{239,123,5}	
\definecolor{ukesec3}{RGB}{104,195,205}	
\definecolor{ukesec4}{RGB}{138,189,36}	
\definecolor{ukesec5}{RGB}{178,34,41}	
\definecolor{tuhh}{RGB}{45,198,214}     
\definecolor{ibidarkBG}{RGB}{227,229,242}   
\definecolor{uke2BG}{RGB}{233,228,225} 	    
\definecolor{uke3BG}{RGB}{230,231,232}	    
\definecolor{ukesec1BG}{RGB}{255,243,190}   
\definecolor{ukesec2BG}{RGB}{254,232,212}   
\definecolor{ukesec3BG}{RGB}{222,241,241}   
\definecolor{ukesec4BG}{RGB}{233,243,222}   
\definecolor{ukesec5BG}{RGB}{244,230,225}   

\pgfplotsset{
  colormap/vik/.style={%
    /pgfplots/colormap={vik}{%
      rgb=(0.001328,0.069836,0.379529)
      rgb=(0.008947,0.163152,0.438691)
      rgb=(0.010254,0.253516,0.49702)
      rgb=(0.03198,0.349543,0.558906)
      rgb=(0.146607,0.458643,0.631289)
      rgb=(0.317174,0.578997,0.71295)
      rgb=(0.503691,0.69791,0.79398)
      rgb=(0.690457,0.811742,0.871059)
      rgb=(0.868874,0.900206,0.914812)
      rgb=(0.933752,0.858522,0.815421)
      rgb=(0.89269,0.743148,0.657086)
      rgb=(0.844258,0.629296,0.505146)
      rgb=(0.798475,0.52238,0.362835)
      rgb=(0.753316,0.419216,0.22819)
      rgb=(0.684653,0.299264,0.095633)
      rgb=(0.562268,0.168171,0.022523)
      rgb=(0.449521,0.078741,0.025203)
      rgb=(0.350423,6.1e-5,0.030499)
    }
  },
  }

\tikzset{
  pics/square/.default={1},
  pics/square/.style = {
    code = {
    \draw[pic actions] (0,0) rectangle (#1,0.6#1);
    }
  }
} 

\begin{document}
\begin{acronym}
  \acro{MPI}{magnetic particle imaging}
  \acro{SM}{system matrix}
  \acro{DF}{drive field}
  \acro{SF}{selection field}
  \acro{DL}{deep learning}
  \acro{PSNR}{peak signal-to-noise ratio}
  \acro{SSIM}{structural similarity index measure}
  \acro{PConvUNet}{PConvUNet}
  \acro{SNR}{signal-to-noise ratio}
  \acro{DCT-F}{multidimensional discrete cosine transform and soft thresholding}
  \acro{GT}{ground truth}
  \acro{FOV}{field of view}
  \acro{CI}{confidence interval}
  \acro{FFP}{field-free point}
  \acro{FFL}{field-free line}
  \acro{FWHM}{full width at half maximum}
\end{acronym}

\articletype{Paper} 

\title{Deep Learning for Restoring MPI System Matrices Using Simulated Training Data}

\author{
Artyom Tsanda$^{1,2,*}$\orcid{0009-0009-7765-4604}, 
Sarah Reiss$^{1,2}$\orcid{0009-0006-4015-1200}, 
Konrad Scheffler$^{1,2}$\orcid{0000-0002-6842-9204}, 
Marija Boberg$^{1,2}$\orcid{0000-0003-3419-7481}, 
and Tobias Knopp$^{1,2,3}$\orcid{0000-0002-1589-8517}}

\affil{$^1$Institute for Biomedical Imaging, Hamburg University of Technology, Hamburg, Germany}

\affil{$^2$Section for Biomedical Imaging, University Medical Center Hamburg-Eppendorf, Hamburg, Germany}

\affil{$^3$Fraunhofer Research Institution for Individualized Medical Technology and Engineering IMTE, Lübeck, Germany}

\affil{$^*$Author to whom any correspondence should be addressed.}

\email{artyom.tsanda@tuhh.de}

\keywords{Magnetic Particle Imaging, System Matrix Recovery, Machine Learning, Image Restoration}

\begin{abstract}
\textit{Objective.}
Magnetic particle imaging reconstructs tracer distributions using a system matrix obtained through time-consuming, noise-prone calibration measurements. Methods for addressing imperfections in measured system matrices increasingly rely on deep neural networks, yet curated training data remain scarce. This study evaluates whether physics-based simulated system matrices can be used to train deep learning models for different system matrix restoration tasks, i.e., denoising, accelerated calibration, upsampling, and inpainting, that generalize to measured data.
\\
\textit{Approach.} A large dataset of system matrices was generated using an equilibrium magnetization model extended with uniaxial anisotropy. The dataset spans particle, scanner, and calibration parameters for 2D and 3D trajectories, and includes background noise injected from empty-frame measurements. For each restoration task, deep learning models were compared with classical non-learning baseline methods. Quantitative performance was evaluated on simulated data using peak signal-to-noise ratio~(PSNR) and structural similarity index measure~(SSIM). For measured data, performance was assessed qualitatively by visual comparison of system matrices and the resulting reconstructions.
\\
\textit{Main results.} The models trained solely on simulated system matrices generalized to measured data across all tasks: for denoising, DnCNN/RDN/SwinIR outperformed \acs{DCT-F} baseline by $>10$ dB PSNR and up to $+0.1$ SSIM on simulations and led to perceptually better reconstructions of real data; for 2D upsampling, SMRnet exceeded bicubic by $\sim 20$~dB PSNR and $\sim 0.08$ SSIM at $\times 2$--$\times 4$ but these gains did not transfer qualitatively to real measurements. For 3D accelerated calibration, SMRnet matched tricubic in noiseless cases and was more robust under noise, and for 3D inpainting, biharmonic inpainting was superior when noise-free but degraded with noise, while a \acs{PConvUNet} maintained quality and yielded less blurry reconstructions.
\\
\textit{Significance.} The demonstrated transferability of deep learning models trained on simulations to real measurements mitigates the data-scarcity problem, which intensifies with model scale. This enables the development of new methods beyond current measurement capabilities and supports pre-training of large models on simulated data.
\end{abstract}

\section{Introduction}

\Ac{MPI} is a preclinical tomographic imaging modality recovering the spatial distribution of superparamagnetic iron oxide nanoparticles \parencite{Gleich2005}. Due to inherent magnetic properties, the particles produce a detectable signal when exposed to an oscillating magnetic field. Spatial encoding in \ac{MPI} is achieved through the superposition of a spatially homogeneous oscillating \ac{DF} and a static gradient field, known as the \ac{SF}. The \ac{SF} defines regions of low magnetic field strength, practically either a \ac{FFP} or a \ac{FFL}, within which particles contribute to the measured signal. The spatial encoding is achieved by moving this region across the \ac{FOV}. MPI provides quantitative, background-free, and real-time imaging capabilities without the use of ionizing radiation \parencite{knopp_magnetic_2017}. Potential clinical applications include cardiovascular imaging \parencite{weizenecker2009three,10.1371/journal.pone.0156899,doi:10.1177/1526602819851202,Mohtashamdolatshahi2020,doi:10.1148/radiol.12120424}, perfusion assessment \parencite{Molwitz_2019,ludewig_magnetic_2022}, and localized magnetic hyperthermia \parencite{doi:10.1021/acsnano.8b00893,healy_clinical_2022}.

The absence of particle-particle interactions makes the \ac{MPI} signal linear with respect to particle concentration. For this reason, measuring the response to a small delta sample of known concentration, moved across a predefined grid, known as the calibration process, yields the forward mapping from concentration distributions to the signal, called the \ac{SM}. The \ac{SM} defines the linear system used for reconstructing unknown concentration maps, thereby playing a decisive role in the quality of the resulting images. Alternatively, the \ac{MPI} signal can be characterized by a point-spread function, leading to the x-space approach \parencite{5728922}.

Acquiring a \ac{SM} is a time-consuming process, it can take about \num{32} hours for a $37 \times 37 \times 37$ grid. Yet, being dependent on scanner, acquisition, and particle parameters, the \ac{SM} needs to be re-measured every time one of them changes. For this reason, accelerating the \ac{SM} measurement is actively being researched. In early work, compressed sensing was applied to recover a full \ac{SM} acquired on an incoherently downsampled grid \parencite{knopp_sparse_2013, weber_reconstruction_2015, 8630849}. Super-resolution methods, primarily based on \ac{DL}, have been applied to a regularly downsampled grid \parencite{10.1007/978-3-030-59713-9_8, kluth_joint_2020, gungor_deep_2021, gungor_transms_2022, shi_progressive_2023, zhang_iterative_2025}. Alternatively, when sampled sufficiently, the calibration size of a measured \ac{SM} can be further increased, resulting in reconstructions with improved perceived image quality. Another downside of the measured \ac{SM} is the presence of noise; although averaging mitigates it, the effect persists at higher frequencies. The problem has been addressed by frequency-domain filters \parencite{weber_denoising_2015} and data compression \parencite{grosser_using_2020}. Lastly, measurements for some positions may be corrupted, often becoming evident only retrospectively. The described imperfections form a set of \ac{SM} restoration problems considered in this paper: denoising, accelerated calibration, upsampling, and inpainting. In the \ac{MPI} community, the term “\ac{SM} recovery” is primarily used for the estimation of missing \ac{SM} values caused by downsampling. In this work, this task is referred to as “accelerated calibration”, and the term “restoration” is used to denote all \ac{SM}-related problems considered here.

Despite the trend towards \ac{DL}-based \ac{SM} restoration methods and the dependency of the \ac{SM} on multiple parameters, the amount of available \ac{SM} data remains limited. The majority of methods rely on the Open~\ac{MPI} dataset, which provides \acp{SM} measured on a preclinical \ac{FFP} scanner (Bruker, Ettlingen)\parencite{KNOPP2020104971}. It includes a few grid sizes ($19 \times 19 \times 19$, $33 \times 33 \times 27$, and $37 \times 37 \times 37$) and two tracer types (Perimag and Synomag-D, Micromod GmbH, Germany). Given its limited diversity and modest size, the dataset may pose challenges for training \ac{DL} models that generalize across different systems and acquisition settings. Consequently, high-quality datasets covering the variability of the parameter space are still in demand for training \ac{DL} models. 

Alternatively, the \ac{SM} can be computed using a physical model \parencite{knopp_model-based_2010, kluth_mathematical_2018}. Accurately describing particle dynamics becomes critical for this approach. For example, taking into account relaxation effects leading to anisotropies by solving the Fokker-Planck equation for the probability density function of magnetic moments. In this case, the task becomes computationally expensive. Recently, the equilibrium model has been extended to take anisotropies into account while maintaining reasonable reconstruction quality \parencite{eq_model_maas}. The primary goal of the described approaches is to substitute the tedious calibration process; however, they are rarely applied in the \ac{DL} context.

In this work, we address the problem of data scarcity in \ac{SM} restoration using \ac{DL} models. Although direct measurement remains the most accurate method for characterizing an \ac{MPI} system, we argue that simulated \ac{SM} data are sufficiently accurate to train \ac{DL}-based restoration models that can subsequently be applied to real observations. To demonstrate this, we generated a training dataset spanning the \ac{SM} parameter space and considered several restoration problems showing the generalization capability of the trained \ac{DL} models. This paper has the following major contributions:
\begin{enumerate}
    \item A method for generating a comprehensive modeled \ac{SM} dataset spanning particle, scanner, and calibration parameters for 2D/3D Lissajous trajectories.
    \item Applications, such as denoising, accelerated calibration, upsampling, and inpainting, showing that \ac{DL} models trained exclusively on simulated \acp{SM} generalize to measured data.
    \item New restoration methods for \ac{SM} denoising and inpainting.
\end{enumerate}

This paper is an extension of the conference abstract \parencite{tsanda_denoising_2025}. Compared to the abstract, we made the following major extensions. First, we expanded the parameter space for \ac{SM} simulation, including 3D trajectories. Second, we improved the quality of denoising by employing new methods. Third, we considered new applications, namely upsampling and inpainting.

In line with the principles of open science, we are making the source code~\parencite{tsanda_2025_17897424} and data~\parencite{tsanda_data_supplement} used in this study publicly available to ensure reproducibility and support future research.

\section{Background}

\subsection{Related Works}

In previous works, simulated data have been primarily used for quantitative evaluations of the proposed methods.  \acp{SM} simulated for the in-house open-sided \ac{FFL} \ac{MPI} scanner \parencite{top_tomographic_2020} were employed to evaluate the proposed \ac{SM} super-resolution method and to perform ablation studies on individual components of the \ac{DL} model \parencite{gungor_transms_2022}. The dataset included \num{100} 2D simulated \acp{SM} with varied SF gradients and particle diameters.
A pre-training strategy using pseudo-labeled low-resolution \ac{SM} data, which incorporates frequency and coil information as tokens in the transformer model, was proposed \parencite{shi_progressive_2023}. The approach allows for more effective fine-tuning on real data later on. The method was validated on a simulated dataset, namely four 3D \acp{SM} were simulated using Langevin model with different SF gradients (\qtylist{0.5;1.0;5}{T.m^{-1}.\mu_0^{-1}}) and particle diameters (\qtylist{25;12.5}{nm}).
Unlike previous approaches, we show that models trained on simulated data can generalize to measured \acp{SM}.

Concurrently, a \ac{DL}-based denoising approach for \acp{SM} was introduced~\parencite{liu_deep_2025}. The authors evaluated several denoising models, including DnCNN~\parencite{zhang2017beyond} and SwinIR~\parencite{Liang20211833} considered in this paper, and proposed a new encoder-decoder architecture with elements of the Swin Transformer~\parencite{liu_swin_2021}. Their models were trained on a single 2D simulated \ac{SM} generated using the Langevin model with added white Gaussian noise, and validated on the Open~\ac{MPI} dataset with synthetic Gaussian noise as well as on an in-house dataset. That study focused on methodological developments for denoising. In contrast, our study investigates the transferability of \ac{DL} models trained on simulated data in general considering a broad range of applications. We employ a more comprehensive simulation framework, vary simulation parameters, and incorporate real background measurements as the noise source. For denoising specifically, we additionally consider 3D data and observe improved qualitative performance for DnCNN and SwinIR on measured phantoms, with noticeably reduced noise in the reconstructions.

\subsection{MPI Signal Model and System Function}

With negligible particle-particle interactions, the \ac{MPI} signal in the frequency domain $u_{l,k} \in \mathbb{C}$ measured in the imaging volume $\Omega \subset \mathbb{R}^3$ from receive coil $l\in\{1,\dots,L\}$ at frequency index $k\in\{0,\dots,K-1\}$ is related to the spatial particle distribution $c : \Omega \to \mathbb{R}$ by
\begin{equation}\label{eq:MPI_signal_model}
u_{l,k} = \int_{\Omega}  \underbrace{\left( -a_k \frac{\mu_0}{T}  \bm{p}_l(\bm{r}) \cdot 
\int_0^T e^{-2\pi i kt/T} \frac{\partial}{\partial t} \overline{\bm{m}} \big(\bm{H}(\bm{r}, t),t) \, dt \right)}_{=\vcentcolon\, s_{l, k}(\bm{r})}c(\bm{r}) \, d\bm r, 
\end{equation}
where 
$s_{l, k} : \Omega \to \mathbb{C}$ is the system function,
$a_k \in \mathbb{C}$ is the transfer function of the analog filter in the signal acquisition chain, 
$\bm{p}_l: \Omega \to \mathbb{R}^3$ is the sensitivity profile of the receive coil $l$, 
$T$ is the acquisition time, 
$\overline{\bm{m}}: \mathbb{R}^3 \times (0, T) \to \mathbb{R}^3$ is the mean magnetic moment depending on the applied magnetic field 
$\bm{H}: \Omega \times (0, T) \to \mathbb{R}^3$,
$\mu_0$ is the vacuum permeability~\parencite{kluth_mathematical_2018}.

To evaluate the system function in \autoref{eq:MPI_signal_model}, the mean magnetic moment must be estimated, which is challenging because the underlying distribution of magnetic moments is unknown. Assuming thermodynamic equilibrium and a static magnetic field, an analytical solution can be derived based on the Langevin function. Albeit being one of the most studied, the Langevin model does not take into account relaxation effects present in real particle systems. A more accurate approach requires solving the Landau-Lifshitz-Gilbert equation using the Fokker-Planck method, which is computationally expensive \parencite{kluth_mathematical_2018}. To avoid these complications, the system function is typically measured using a small sample $\bm \Delta: \Omega \to \{0, 1\}$ at a position $\bm{r}_n \in \Omega$, $n\in\{1,\dots,N\}$, filled with a fixed particle concentration $c_0$:
\begin{equation}
u^n_{l,k}
= \int_{\Omega} s_{l,k}(\bm{r}) c_0 \bm \Delta(\bm{r} - \bm{r}_n) \, d\bm r
\approx c_0 s_{l,k}(\bm{r}_n) 
\underbrace{\int_{\Omega} \bm \Delta(\bm r) \, d\bm r}_{=: V_{\Delta}}.
\end{equation}
The measured discrete system function, hereafter referred to as the system matrix, is then formed by moving the sample on a discrete Cartesian grid and repeating the measurement. 
The resulting matrix $\bm S \in \mathbb{C}^{M \times N}$ maps $N$ spatial locations to $M=KL$ frequencies from receive coils.

Recently, an extension to the equilibrium model has been proposed to account for anisotropies in magnetization dynamics \parencite{eq_model_maas}. It adds an anisotropy term to the partition function:
\begin{equation}\label{eq:partition_func_anisotropy}
\mathcal{Z}(\beta\bm{H}; \mathbb{O}) = \int_{\mathbb{S}^2} e^{\beta\bm{H}^\mathrm{T}\bm{m} + \alpha_K(\bm r) (\bm{n}(\bm r)^\mathrm{T}\bm{m})^2} \, d\bm{m},
\end{equation}
with
\begin{equation}\label{eq:beta}
\beta \vcentcolon= \frac{\mu_0 m_0}{k_{\text{B}} T_{\text{P}}}.
\end{equation}
Here, $k_{\text{B}}$ is the Boltzmann constant, $T_{\text{P}}$ is the particle temperature, $m_0 \in \mathbb{R}^+$ is the magnitude of the magnetic moment of a single particle, $\alpha_K: \mathbb{R}^3 \to \mathbb{R}$ and $\bm{n}: \mathbb{R}^3 \to \mathbb{S}^2$ are, respectively, the spatially varying anisotropy constant and the uniaxial easy‑axis direction; together they form the observable parameter set $\mathbb{O}$. 
The model describes immobilized particles, for which both the easy axis and the anisotropy constant are fixed, and fluid particles, for which the easy axis is oriented along the \ac{SF} and the anisotropy depends on the spatial modulation parameter $q \in \mathbb{R}_{+}$~\parencite{eq_model_maas}.
The mean magnetic moment can then be evaluated using
\begin{equation}\label{eq:mean_magn_moment_anisotropy}
\overline{\bm{m}} \big(\bm{H}) =\frac{m_0}{\beta} \nabla_{\bm{H}} \ln (\mathcal{Z}(\beta\bm{H}); \mathbb{O}).
\end{equation}
In this case, \autoref{eq:mean_magn_moment_anisotropy} cannot be solved in closed form but can be approximated using truncated series of Bessel functions. The resulting \ac{SM} provides accurate enough estimations for a SM to reconstruct real data while being computationally less demanding than, for example, the Landau-Lifshitz-Gilbert equation approach. For this reason, we select this model in this paper.

\subsection{Reconstruction}

Having a system matrix $\bm S \in \mathbb{C}^{M \times N}$ and a measurement $\bm{u}_{\text{meas}} \in \mathbb{C}^{M}$, we reconstruct the unknown concentration vector $\bm{c} \in \mathbb{R}^{N}_{+}$ by solving the Tikhonov regularized least squares problem:
\begin{equation}\label{eq:recon_problem}
\underset{\bm{c} \in \mathbb{R}_+^N}{\operatorname{argmin}}
\ \left\| \bm{W}\tilde{\bm{P}}_{\Theta}(\bm S)\bm{c}
- \bm{W}\bm{P}_{\Theta}(\bm{u}_{\text{meas}})\right\|_2^2
+ \lambda \, \|\bm{c}\|_2^2,
\end{equation}
where $\bm{P}_{\Theta}: \mathbb{C}^{KL} \rightarrow \mathbb{C}^{K_\Theta L}$ and $\tilde{\bm{P}}_{\Theta}: \mathbb{C}^{KL \times N} \rightarrow \mathbb{C}^{K_\Theta L \times N}$ are the frequency selection  based on \ac{SNR} threshold $\Theta$ and resulting in $K_\Theta$ frequencies for the measurement and the \ac{SM}, respectively, $\bm{W} \in \mathbb{R}_{+}^{K_\Theta L \times K_\Theta L}$ is the weighting based on the $L^2$ norm of the \ac{SM} rows, $\lambda$ is the regularization parameter. In this work, the regularization parameter is not further scaled but reported as is. The optimization problem \autoref{eq:recon_problem} is solved using the Kaczmarz method with a fixed number of iterations $n_{\text{iter}} \in \mathbb{N}$ \parencite{knopp2010weighted}.

\section{Method}

\subsection{Data Generation}\label{sec:data_generation}

\acp{SM} in \ac{MPI} depend on many parameters. To structure the simulation process, we divide them into three groups: particle, scanner, and calibration parameters. The summary for each of those can be found in \autoref{table:simulation-parameters}.

As we employ the equilibrium model with anisotropy from \autoref{eq:partition_func_anisotropy} to calculate the dynamics of the mean magnetic moment, the particle parameters are $\beta$, $\alpha_K$, $\bm{n}$, and $q$. 
The magnetic moment of a single particle $m_0$ depends on the particle parameter $d_{\text{P}}$ via:
\begin{equation}\label{eq:m0}
    m_0 = \frac{\pi d_{\text{P}}^3 M_{S}}{6},
\end{equation}
where $M_{S}$ is the saturation magnetization of the particle core, taken fixed at \qty{474000}{J.m^3.T^{-1}} \parencite{deissler_relax}.
We vary the particle diameter $d_{\text{P}}$, which changes the magnetic moment of a single particle $m_0$ and thus the parameter $\beta$ (see  \autoref{eq:beta}).
We select the same range of particle diameters as used by \cite{eq_model_maas}, namely \qtyrange{15}{25}{nm}. The cubed diameter is then sampled uniformly between the corresponding cubed values. The temperature is taken fixed at \qty{293}{K}. The anisotropy constant $K^{\text{anis}}$ is drawn from a log-uniform distribution over the interval $[10^3, 10^4]\,\unit{J.m^{-3}}$. To augment the resulting dataset, we consider both immobilized and fluid particles. Each particle type is selected with probability \num{0.5} for every simulated \ac{SM}. When immobilized, the direction of the easy axis is sampled uniformly on the unit sphere.
When fluid, the spatial modulation parameter $q$ is sampled uniformly from \qtyrange{0.3}{1.3}, where the range is selected heuristically.

In this paper, we limit data generation to \ac{FFP} scanners with Lissajous trajectories. In this case, the SF is parametrized with 
three spatial gradients $G_x, G_y, G_z$. We select the first two components along $x$ and $y$ uniformly from \qtyrange{0.1}{1.5}{T.m^{-1}.\mu_0^{-1}}. The gradient along $z$ is then calculated as the negative sum of the first two to fulfill Gauss's law for magnetism. The range of DF amplitudes $A_i,~i\in\{x,y,z\}$, is set to
\qtyrange{5}{14}{mT.\mu_0^{-1}}. Both ranges for \ac{SF} gradients and \ac{DF} amplitudes were selected according to hardware specifications of several existing \ac{MPI} systems: a preclinical scanner (Bruker, Ettlingen) \parencite{KNOPP2020104971}, a human-sized scanner \parencite{thieben_system_2024}, and an open-sided scanner \parencite{top_tomographic_2020}. 
We consider 2D and 3D Lissajous trajectories. \ac{DF} frequencies $f_x$, $f_y$, $f_z$ remain fixed at $2.5/102,\,2.5/96,\,2.5/99\,\mathrm{MHz}$ respectively.

In the final subset of parameters, we specify the calibration. The calibration \ac{FOV} is linked to the \ac{DF} \ac{FOV}: along each axis $i\in\{x, y, z\}$, $\text{FOV}^{\text{calib}}_{i}$ is drawn from $\text{FOV}^{\text{DF}}_{i}\cdot \mathcal{U}(1,2)$ where $\text{FOV}^{\text{DF}}_i = 2A_i/\,|G_i\,|$. Here, $\mathcal{U}(a, b)$ denotes the uniform distribution on the interval $[a, b]$. To fully cover the resulting \ac{SM}, the center of the calibration \ac{FOV} is sampled uniformly within the interval $[-m_i, m_i]$ defined by the margins $m_i = (\text{FOV}^{\text{calib}}_{i} - \text{FOV}^{\text{DF}}_{i})/2$.
We sample the calibration size (pixels per axis) based on the 1D \ac{MPI} convolution kernel’s \ac{FWHM} \parencite{knopp_limitations_2010}:
\begin{equation}
    N^{\text{calib}}_i
    = \left\lfloor \left( \frac{\text{FOV}^{\text{calib}}_i}{R^{\text{FWHM}}_i}\cdot \mathcal{U}(6.24, 8.32) \right) \right\rceil,
    \qquad
    R^{\text{FWHM}}_i=\frac{4.16}{\beta \,|G_i\,|},
\end{equation}
where $G_i$ is the \ac{SF} gradient along each axis.

\begin{figure}
 \centering
        \includegraphics[width=1.0\textwidth]{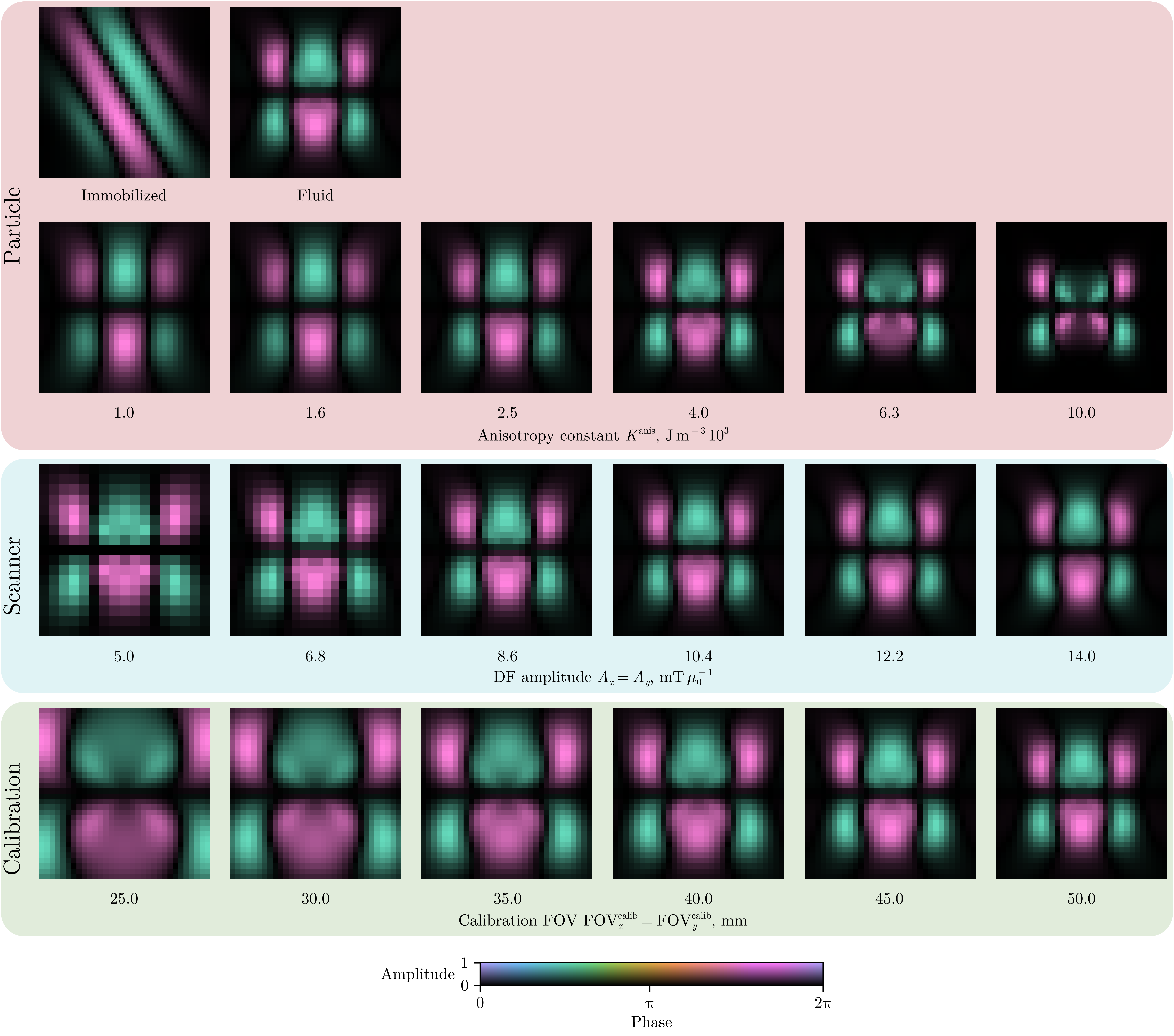}
 \caption{
 Simulated 2D \acp{SM} at the frequency $f=99.5$\,\unit{kHz} illustrating variations across parameter groups: anisotropy constant (particle), \ac{DF} amplitude (scanner), and calibration \ac{FOV} (calibration). Unless stated otherwise, parameters are fixed to $d_{\text{P}}=20\,\unit{nm}$, $K^{\text{anis}}=3.2\,\unit{J.m^{-3}.10^3}$, $q=1$, $\bm{n}=[\cos(-3\pi/4),\,\sin(-3\pi/4)]$, $A_x=A_y=10\,\unit{mT.\mu_0^{-1}}$, $G_x=G_y=0.8\,\unit{T.m^{-1}.\mu_0^{-1}}$, $\mathrm{FOV}^{\mathrm{calib}}_x=\mathrm{FOV}^{\mathrm{calib}}_y=4A_x/G_x$, and $N^{\mathrm{calib}}_x=N^{\mathrm{calib}}_y=7\cdot \mathrm{FOV}^{\mathrm{calib}}_x/R^{\mathrm{FWHM}}_x$.
 }
\label{fig:sm-examples}
\end{figure}

Following the described sampling of simulation parameters we generate 1000/300/300 2D and 50/15/15 3D \acp{SM} for the training, validation, and test sets, respectively. A few examples are shown in \autoref{fig:sm-examples}.

\begin{table}
\centering
\caption{Simulation parameters used for data generation.}
\begin{tabularx}{1.0\linewidth}{llX}
\toprule
\textbf{Parameter} & & \textbf{Value} \\
\midrule
\multicolumn{3}{c}{\textbf{Particle}} \\
\midrule
Temperature & $T_{\text{P}}$ & \qty{293}{K} \\
Sat. magnetization & $M_{S}$ & \qty{474000}{J.m^3.T^{-1}} \\
Particle core diameter & $d_{\text{P}}$ & $\mathcal{U}(15^3, 25^3)^{1/3}$\,nm \\
Anisotropy constant & $K^{\text{anis}}$ & $10^{\mathcal{U}(3,4)}\,\unit{J.m^{-3}}$ \\
Fluid/Immobilized & & $\text{Bernoulli}(p=0.5)$ \\
Spatial modulation (fluid) & $q$ & $\mathcal{U}(0.3,1.3)$\\
Easy-axis direction (immob.)& $\bm{n}$ & $\mathcal{U}(\mathbb{S}^2)$ \\
\midrule
\multicolumn{3}{c}{\textbf{Scanner}} \\
\midrule
\ac{DF} amplitudes & $[A_x,\,A_y,\,A_z]$ & $[\mathcal{U}(5,14),\,\mathcal{U}(5,14),\,\mathcal{U}(5,14)]$\,\unit{mT.\mu_0^{-1}} \\
\ac{DF} frequencies & $[f_x,\,f_y,\,f_z]$ & $[2.5/102,\,2.5/96,\,2.5/99]$\,\unit{MHz} \\
\ac{SF} gradient & $[G_x,\,G_y,\,G_z]$ & $[\mathcal{U}(0.1,1.5),\,\mathcal{U}(0.1,1.5),\,-(G_x+G_y)]$\,\unit{T.m^{-1}.\mu_0^{-1}} \\
Trajectory & & 2D/3D Lissajous \\
\ac{DF} \ac{FOV} per axis & $\mathrm{FOV}^{\mathrm{DF}}_i$ & $2A_i/|G_i|$\,\unit{mm} \\
\midrule
\multicolumn{3}{c}{\textbf{Calibration}} \\
\midrule
Calibration \ac{FOV} per axis & $\mathrm{FOV}^{\mathrm{calib}}_i$ & $\mathrm{FOV}^{\mathrm{DF}}_i \cdot \mathcal{U}(1,2)$\,\unit{mm} \\
Margins per axis & $ m_i $ & $(\text{FOV}^{\text{calib}}_{i} - \text{FOV}^{\text{DF}}_{i})/2$\,\unit{mm} \\
Calibration center per axis & & $\mathcal{U}(-m_i, m_i)$\,\unit{mm}  \\
Calibration size (voxels/axis) & $N^{\text{calib}}_i$ & $\lfloor \text{FOV}^{\text{calib}}_i/R^{\text{FWHM}}_i\cdot \mathcal{U}(6.24, 8.32) \rceil$ \\
\bottomrule
\end{tabularx}
\label{table:simulation-parameters}
\end{table}

\subsection{Restoration problems}

We consider four degradation problems: denoising, accelerated calibration, upsampling, and inpainting. Having a simulated ground-truth \ac{SM} $\bm S_{\text{GT}}$, we corrupt it according to the problem. The corruption model can be generally described by the following equation
\begin{equation}\label{eq:corruption_models}
    \bm S^{p}_{\text{corrupt}} = \mathcal{A}(\bm S_{\text{GT}}) + \bm N,
\end{equation}
where $\bm N \in \mathbb{C}^{M \times N}$ is a noise matrix and $\mathcal{A}: \mathbb{C}^{M \times N} \to \mathbb{C}^{M \times N}$ is a linear transform specific to each problem: (i) identity for denoising; (ii)  regular downsampling, selecting each $n$-th pixel along each dimension, for accelerated calibration and upsampling; (iii) masking pixels corresponding to valid calibration measurements for inpainting.

Practically, in addition to the particle-induced signal $\bm u$ described in \autoref{eq:MPI_signal_model}, the measured voltage $\bm u_{\text{meas}}$ contains the following components:
\begin{equation}\label{eq:noise_components}
    \bm u_{\text{meas}} = \bm u + \bm u^{\text{BG}} + \bm u^{\text{N}} + \bm u^{\text{D}},
\end{equation}
where $\bm u^{\text{BG}} \in \mathbb{C}^{M}$ is the systematic background signal, $\bm u^{\text{N}} \in \mathbb{C}^{M}$ is thermal noise, $\bm u^{\text{D}} \in \mathbb{C}^{M}$ corresponds to abrupt signal changes or distortions, which may be caused by hardware imperfections or external factors \parencite{paysen_characterization_2020}. In this work, we consider all terms except the particle signal to be noise. The drifts inside the background signals and the abrupt component 
are typically non‑stationary in time and therefore affect different calibration measurements to varying degrees. Since a \ac{SM} is measured consecutively for multiple positions, these temporal fluctuations result in spatially structured artifacts and inconsistent noise levels. 
As these effects are challenging to model, we take a sequence of background measurements as a source of noise. In particular, we measured \num{7801200} and \num{500000} frames for 2D and 3D Lissajous trajectories respectively with the preclinical \ac{FFP} scanner (Bruker, Ettlingen). To sample noise for a particular \ac{SM}, we select a sub-sequence of frames and reshape them into the required size selecting the corresponding channel and frequency.

\paragraph{Denoising:}

As a baseline method for denoising, we use previously proposed frequency-domain filtering based on \ac{DCT-F} \parencite{weber_denoising_2015}. For each complex frequency component (per frequency and channel), a discrete cosine transform is applied along the spatial axes, and soft thresholding is performed by shrinking the magnitude by $\omega \sigma$ while preserving its phase; the denoised image is then obtained via the inverse discrete cosine transform. We set $\omega = 2.75$. The method is non-blind and requires the noise standard deviation $\sigma$. For simulated data, we used the ground-truth values, whereas for real measurements we estimated $\sigma$ from background noise measurements and heuristically adjusted it by a coefficient (0.3). We consider three deep learning methods for the denoising problem: DnCNN \parencite{zhang2017beyond}, RDN \parencite{zhang_residual_2021}, and SwinIR~\parencite{Liang20211833}. DnCNN represents a sequence of convolution, normalization, and activation layers. In our prior study \parencite{tsanda_denoising_2025}, we provided first denoising results with this method. In this work, we extend the method to 3D trajectories. RDN is composed of residual dense blocks and was used in \ac{MPI} before for the Plug-and-Play reconstruction \parencite{askin_pp-mpi_2022, tsanda_pnp_2024}. Finally, SwinIR employs a parameter-effective transformer architecture with windowed self-attention layers and shows state-of-the-art denoising performance for natural images. However, due to hardware limitations, we do not apply SwinIR to 3D problems.

\paragraph{Accelerated calibration:}

The accelerated calibration task aims to reconstruct a fully sampled \ac{SM} from a regularly downsampled 3D calibration grid. Regular downsampling induces aliasing and information loss, thereby degrading reconstruction quality. In contrast to spatial super-resolution, accelerated calibration must additionally account for measurement noise. We therefore formulate it as a joint completion-and-denoising problem. As a baseline method, we use tricubic spline interpolation applied independently across channels and frequencies. As a data-driven approach, we employ SMRnet \parencite{10.1007/978-3-030-59713-9_8}, which consists of stacked residual dense blocks and a nearest-neighbor upsampling module followed by convolutional layers.

\paragraph{Upsampling:}

Although the formal degradation operator is identical to that of accelerated calibration, we treat upsampling separately due to its distinct objective and noise characteristics. In upsampling, an already complete and adequately sampled \ac{SM} is available and sufficient for reconstruction. The goal is perceptual refinement by increasing the calibration grid resolution beyond the physically achievable optimum. Methods are applied as a post-processing step after denoising. The problem is an example where measured \ac{GT} is not available for training. We study this problem in 2D and generate an additional dataset with larger calibration grids while keeping all other simulation parameters fixed. We evaluate bicubic interpolation as a baseline method and a 2D version of SMRnet.

\paragraph{Inpainting:}

During the \ac{SM} calibration process, individual measurements at different spatial positions can be corrupted by external interference or internal system instabilities. Because a full calibration can last \num{32} hours or more, the risk of corruption is substantial. To date, entire calibration scans are typically discarded and repeated. To mitigate this, we instead propose non-blind inpainting for 3D \acp{SM}, using known masks of invalid pixels provided to the method. Because the artifacts correspond to sequences of failed measurements at consecutive robot positions, we generate masks by first flattening the calibration volume into a one‑dimensional index space, selecting a single contiguous block of missing indices, and then mapping this pattern back to three dimensions. To emulate different traversal patterns (raster or meandering), the order of elements in the last dimension is reversed with probability 0.5 for each mask. In addition, the spatial dimensions are randomly permuted before mask generation, and the permutation is inverted afterwards. The size of the mask is controlled by the ratio with respect to the number of elements in the original \ac{SM}. We also combine multiple masks generated independently into one for training. See the original code for further details on mask generation \parencite{tsanda_2025_17897424}. Since we apply methods independently to each frequency component, a new mask is randomly generated for each. As a baseline method, image inpainting is performed by solving biharmonic partial differential equations that extend image values smoothly into missing regions \parencite{biharmonic1, biharmonic2}. Alternatively, we employ a U-Net architecture with partial convolutions adapted to 3D \parencite{Liu201889}.

\subsection{Training}

All models were trained on complex-valued frequency components of \acp{SM} from the training set. For each component, the \ac{GT} image was amplitude-normalized to unit maximum, rotated by a single random angle on the complex plane, and randomly scaled. Having a pre-processed \ac{GT}, a task-specific degradation operator $\mathcal{A}$ was applied, and measured background noise was added after scaling to a randomly sampled standard deviation. The degraded sample was then re-normalized to the maximum amplitude of the corrupted image. These steps emulate the varying \ac{SNR} in \ac{MPI} and improve robustness to global phase shifts introduced by the acquisition chain’s transfer function (see \autoref{eq:MPI_signal_model}). To handle varying calibration sizes, each image was zero‑padded to a fixed patch size with randomly sampled padding on all sides, resulting in a varying offset within the patch. Finally, complex-valued frequency components were transformed into two-channel real-valued arrays corresponding to the real and imaginary parts.

For denoising, accelerated calibration, and upsampling, the objective was the L1 loss. For inpainting, a weighted L1 loss was used with separate weights for known and missing pixels. In all cases, the loss was computed only on valid (non-padded) pixels of the original sampling grid. Optimization used Adam with a multi-step learning rate schedule \parencite{2015-kingma}. We used mixed precision~\parencite{MicikeviciusNAD18} and applied exponential moving averaging of the model weights during training. All models were trained on a single NVIDIA H100 GPU, except SwinIR, which required two GPUs. Further details can be found in the original code \parencite{tsanda_2025_17897424}.

\subsection{Evaluation}

We evaluate the trained models on both simulated and real \acp{SM}. For simulated data, we fix the \ac{SNR} by normalizing the magnitude of the ground-truth frequency components to their maximum and scaling the additive noise to a selected standard deviation. At inference, inputs are preprocessed as during training: scaled to unit magnitude, padded to the required patch size, and converted to two-channel arrays (real and imaginary parts). After prediction, these transformations are inverted.

For simulated data, per-frequency errors are quantified using \ac{PSNR} and \ac{SSIM} \parencite{1284395} and then averaged over the test set. We report mean values with 95\% \acp{CI}. The \ac{CI} is computed using a normal approximation, namely for 95\% confidence level $\bar{x} \pm 1.96 \cdot \sigma / \sqrt{n}$, where $\bar{x}$ is the sample mean, $\sigma$ is the sample standard deviation computed with Bessel's correction, and $n$ is the sample size.

To assess the generalization to real data, we consider several measurements acquired on a preclinical \ac{FFP} scanner (Bruker, Ettlingen). 
Unless stated otherwise, all real measurements were acquired using Lissajous \ac{DF} sequences with \ac{DF} amplitudes $A_x = A_y = A_z = 12\,\unit{mT.\mu_0^{-1}}$ 
(with $A_z$ applicable only for 3D trajectories) and \ac{DF} frequencies $f_x = 2.5/102\,\unit{MHz}$, $f_y = 2.5/96\,\unit{MHz}$, and, 
for 3D trajectories, $f_z = 2.5/99\,\unit{MHz}$. Data were acquired over three receive channels. For 2D datasets, 
\num{1633} frequency components were retained; for 3D datasets -- \num{53857} components. Unless specified, the tracer was Perimag (Micromod GmbH, Germany), and 
\ac{SF} gradients were $G_x = G_y = -1$\,\unit{T.m^{-1}.\mu_0^{-1}} and $G_z = 2$\,\unit{T.m^{-1}.\mu_0^{-1}}. 
The reconstruction parameters were chosen individually for each measurement and then kept fixed across all methods applied to that measurement.

\begin{figure}
    \centering
    \input{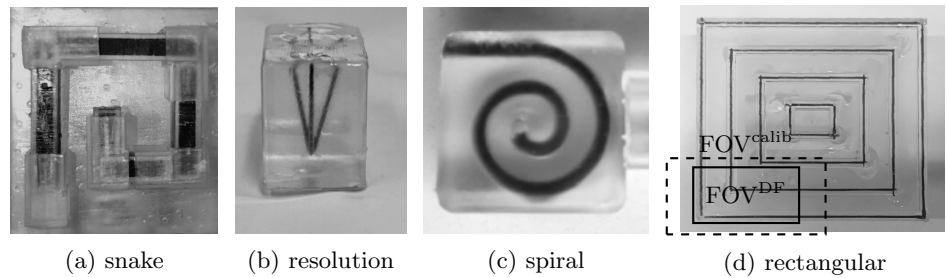}
    \caption{Photographs of the phantoms used for evaluation: (a) snake (denoising); (b) resolution (accelerated calibration); (c) spiral (upsampling); and (d) rectangular (inpainting).}
    \label{fig:phantoms}
\end{figure}

\paragraph{Denoising:}
To evaluate denoising methods, the snake phantom (see \autoref{fig:phantoms-a}) was measured using a 2D Lissajous sequence in the $xy$-plane \parencite{eq_model_maas}. The \ac{SM} was acquired on a \numproduct{17 x 15} Cartesian grid covering a \qtyproduct{34 x 30}{\mm}~\ac{FOV}. At each position the sequence was repeated multiple times. The phantom comprised five rods with \qtyproduct{2.5 x 2.5}{\mm} cross-sections and lengths of \qtylist{20;17.5;15;8.75;5}{\mm} arranged in a meandering (“snake”) pattern. The phantom measurement was averaged \num{1000} times. \ac{SM} measurements were averaged \num{10} and \num{330} times. The 330-average \ac{SM} serves as a reference for denoising. The reconstruction used the first two receive channels with \ac{SNR} threshold of \num{1.5} applied to select frequency components, and the inverse problem was solved using \num{1000} Kaczmarz iterations with Tikhonov regularization ($\lambda=0.3$).

\paragraph{Accelerated calibration:}
For accelerated calibration, the resolution phantom (see \autoref{fig:phantoms-b}) was measured using a 3D Lissajous sequence \parencite{KNOPP2020104971}. The \ac{SM} was acquired on a \numproduct{19 x 19 x 19} Cartesian grid covering a \qtyproduct{24 x 24 x 12}{\mm}~\ac{FOV}. The phantom consists of five tubes originating from a common point and extending at different angles (two at approximately \ang{20} and \ang{30} in the $xy$-plane, two at approximately \ang{10} and \ang{15} in the $yz$-plane, plus one along the $y$-axis). Both the calibration and the phantom measurements were averaged \num{1000} times. To evaluate the proposed methods, the acquired \ac{SM} was downsampled along each dimension by the factor of two starting from the first index which resulted in a \numproduct{10 x 10 x 10} Cartesian grid. Similarly to the original reconstruction, we used frequency selection with \ac{SNR} threshold of \num{3.0}, and \num{3} Kaczmarz iterations with Tikhonov regularization ($\lambda=0.001$).

\paragraph{Upsampling:}
For upsampling, the spiral phantom (see \autoref{fig:phantoms-c}) was measured using a 2D Lissajous sequence in the $xy$-plane \parencite{Mohn_2024}. The \ac{SM} was acquired on a \numproduct{29 x 29} Cartesian grid covering a \qtyproduct{29 x 29}{\mm}~\ac{FOV}. The phantom comprised a spiral with two full windings. The phantom measurement was averaged \num{500} times; \ac{SM} measurements were averaged \num{1000} times. The data were filtered with an \ac{SNR} threshold of \num{1.5} and reconstructed with \num{1000} Kaczmarz iterations with Tikhonov regularization ($\lambda=0.4$).

\paragraph{Inpainting:}
For inpainting, one \ac{SM} corresponding to a corrupted patch from 3D multi-patch measurements of a nested rectangular phantom (see \autoref{fig:phantoms-d}) was used \parencite{8490900}. This experiment employed $G_x = G_y = -0.75$\,\unit{T.m^{-1
}.\mu_0^{-1}} and $G_z = 1.5$\,\unit{T.m^{-1
}.\mu_0^{-1}} and used ferucarbotran (Resovist, I’rom Pharmaceuticals, Tokyo, Japan) as tracer. The \ac{SM} was acquired on a \numproduct{25 x 21 x 27} Cartesian grid covering a \qtyproduct{50 x 42 x 27}{\mm}~\ac{FOV}. The phantom consists of four square-shaped nested tubes in the $xz$-plane, the lower left part of which lies within the \ac{FOV}. The inpainting mask was generated manually to cover corrupted calibration positions. The \ac{SM} and phantom measurements were averaged \num{50} and \num{100} times, respectively. We reconstructed the selected patch according to \autoref{eq:recon_problem} with an \ac{SNR} threshold of \num{10} and \num{10} iterations of the Kaczmarz method with Tikhonov regularization ($\lambda=10^{-5}$).

\subsection{Ablation study: impact of simulation model on transferability to real data}

Generally, the accuracy of the simulation model should directly impact 
the transferability of the trained \ac{DL} models to real data. 
In this study, we select the equilibrium model with anisotropy to 
balance between the computational complexity and the accuracy of the resulting \ac{SM}. 
To demonstrate the importance of the simulation model, we perform an ablation study for the denoising problem in 2D. 
Using the same parameters for particle, scanner, and calibration groups, we generate a dataset using the equilibrium Langevin model.
In this case, the parameters related to anisotropy are ignored.
With the same hyperparameters, we train RDN on the dataset generated with 
the Langevin model, referred to as RDN-Langevin.
For quantitative analysis we employ the test set of \acp{SM} generated 
using the equilibrium model with anisotropy excluding immobilized particles. 
For qualitative analysis, we apply the trained model to the measured \ac{SM} corresponding to the snake phantom.

\section{Results}

\subsection{Denoising}

\autoref{table:results_denoising} summarizes the quantitative results for the denoising task on simulated data with varying noise levels. Compared to the classical \acs{DCT-F} method, the learning-based approaches achieve substantially higher performance across all noise levels, with PSNR improvements exceeding \num{10} dB and SSIM gains of up to \num{0.15} in most cases. For 2D Lissajous trajectories, the SwinIR model consistently achieves the highest \ac{PSNR} and \ac{SSIM} values at every noise level, followed by RDN and DnCNN. This ranking mirrors observations reported on natural-image benchmarks, indicating the transferability of these models to the \ac{MPI} domain. For 3D Lissajous trajectories, RDN performs best in the absence of a 3D SwinIR counterpart, achieving quality levels comparable to those of the 2D setting.

\begin{table}
\caption{Denoising results on simulated SM data with varying noise levels ($\sigma$). Mean $\pm$ 95\% \ac{CI} of \ac{PSNR}/\ac{SSIM} are reported for \ac{DCT-F}, DnCNN, RDN, and SwinIR methods on 2D and 3D Lissajous trajectories.}
\centering
\begin{tabular}{lccr@{ $\pm$ }lr@{ $\pm$ }lr@{ $\pm$ }lr@{ $\pm$ }l}
\toprule
 & $\sigma$ & Metric & \multicolumn{2}{c}{DCT-F} & \multicolumn{2}{c}{DnCNN} & \multicolumn{2}{c}{RDN} & \multicolumn{2}{c}{SwinIR} \\
\midrule
\multicolumn{11}{c}{2D} \\
\midrule
\multirow{2}{*}{} & \multirow{2}{*}{0.06} & PSNR & $25.0$ & $0.1$ & $38.9$ & $0.3$ & $40.6$ & $0.3$ & $41.1$ & $0.3$ \\
 &  & SSIM & $0.841$ & $0.009$ & $0.989$ & $0.001$ & $0.992$ & $0.001$ & $0.993$ & $0.001$ \\
\addlinespace[0.3em]
\multirow{2}{*}{} & \multirow{2}{*}{0.10} & PSNR & $21.9$ & $0.1$ & $35.7$ & $0.3$ & $37.5$ & $0.3$ & $37.8$ & $0.3$ \\
 &  & SSIM & $0.76$ & $0.01$ & $0.978$ & $0.003$ & $0.986$ & $0.002$ & $0.986$ & $0.002$ \\
\addlinespace[0.3em]
\multirow{2}{*}{} & \multirow{2}{*}{0.20} & PSNR & $18.1$ & $0.1$ & $31.1$ & $0.3$ & $33.1$ & $0.3$ & $33.3$ & $0.3$ \\
 &  & SSIM & $0.57$ & $0.01$ & $0.947$ & $0.005$ & $0.966$ & $0.004$ & $0.968$ & $0.003$ \\
\addlinespace[0.3em]
\multirow{2}{*}{} & \multirow{2}{*}{0.30} & PSNR & $16.0$ & $0.1$ & $28.1$ & $0.3$ & $30.4$ & $0.3$ & $30.6$ & $0.3$ \\
 &  & SSIM & $0.42$ & $0.01$ & $0.904$ & $0.007$ & $0.942$ & $0.005$ & $0.946$ & $0.004$ \\
\addlinespace[0.3em]
\midrule
\multicolumn{11}{c}{3D} \\
\midrule
\multirow{2}{*}{} & \multirow{2}{*}{0.06} & PSNR & $28.01$ & $0.08$ & $37.23$ & $0.09$ & $38.30$ & $0.09$ & \multicolumn{2}{c}{---} \\
 &  & SSIM & $0.914$ & $0.003$ & $0.9876$ & $0.0008$ & $0.9898$ & $0.0008$ & \multicolumn{2}{c}{---} \\
\addlinespace[0.3em]
\multirow{2}{*}{} & \multirow{2}{*}{0.10} & PSNR & $25.23$ & $0.08$ & $34.52$ & $0.09$ & $35.79$ & $0.09$ & \multicolumn{2}{c}{---} \\
 &  & SSIM & $0.845$ & $0.004$ & $0.978$ & $0.001$ & $0.9827$ & $0.0010$ & \multicolumn{2}{c}{---} \\
\addlinespace[0.3em]
\multirow{2}{*}{} & \multirow{2}{*}{0.20} & PSNR & $21.74$ & $0.09$ & $30.49$ & $0.09$ & $32.10$ & $0.09$ & \multicolumn{2}{c}{---} \\
 &  & SSIM & $0.692$ & $0.006$ & $0.947$ & $0.002$ & $0.961$ & $0.002$ & \multicolumn{2}{c}{---} \\
\addlinespace[0.3em]
\multirow{2}{*}{} & \multirow{2}{*}{0.30} & PSNR & $19.77$ & $0.09$ & $27.83$ & $0.09$ & $29.58$ & $0.10$ & \multicolumn{2}{c}{---} \\
 &  & SSIM & $0.558$ & $0.007$ & $0.906$ & $0.003$ & $0.930$ & $0.003$ & \multicolumn{2}{c}{---} \\
\addlinespace[0.3em]
\bottomrule
\end{tabular}
\label{table:results_denoising}
\end{table}

\begin{figure}
 \centering
        \includegraphics[width=1.0\textwidth]{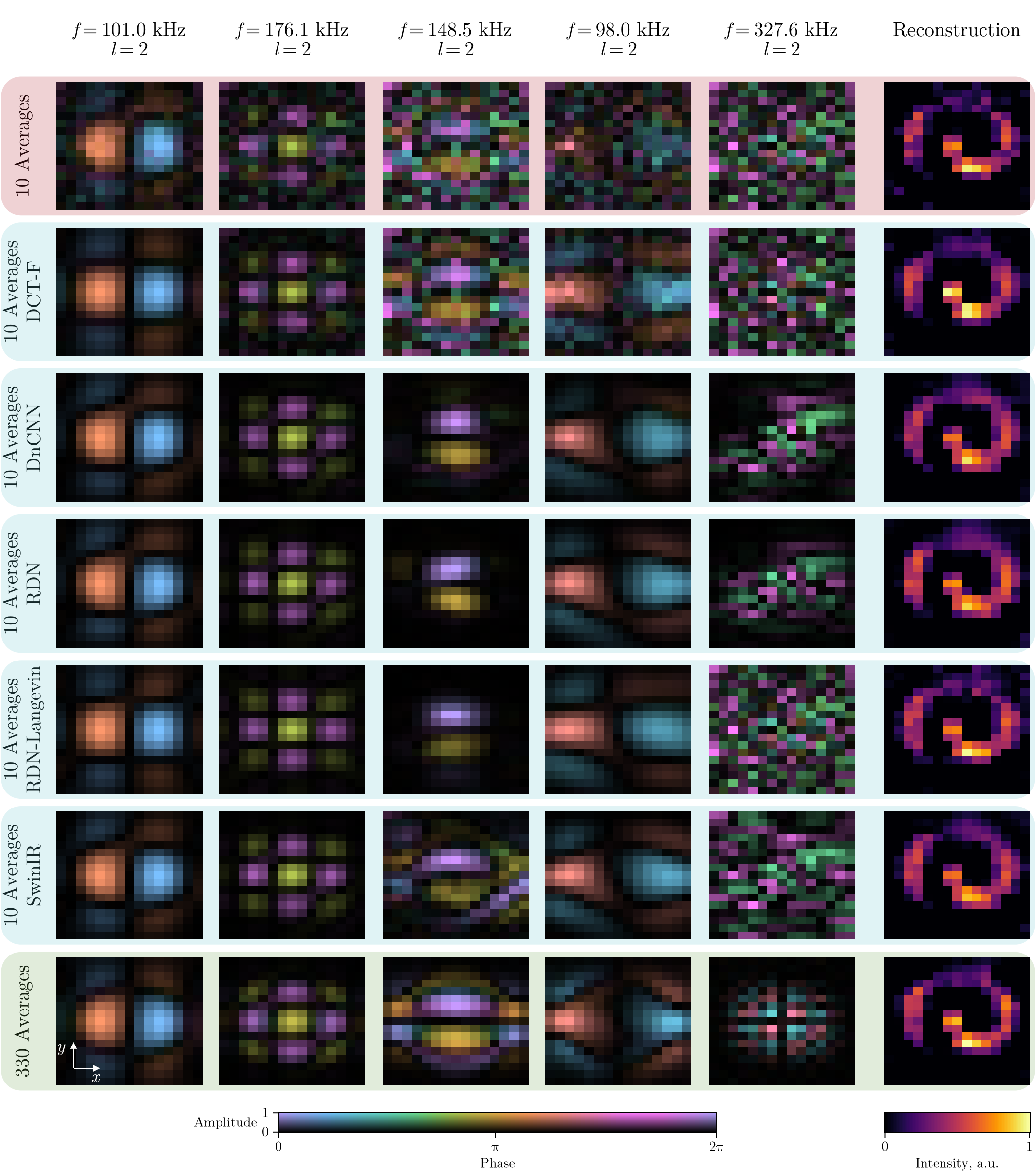}
 \caption{Comparison of denoising methods (\acs{DCT-F}, DnCNN, RDN, RDN-Langevin, and SwinIR) on a measured 2D \ac{SM} corresponding to the snake phantom. Selected frequency components from receive coil $l=2$ and corresponding reconstructions are shown in the $xy$-plane.}
\label{fig:sm-denoising-real}
\end{figure}

The results on real data are presented in \autoref{fig:sm-denoising-real}. In contrast to the simulated test data, the \ac{SNR} of real \acp{SM} varies across frequencies. For frequency components with higher \ac{SNR}, all methods, including the classical \acs{DCT-F} approach, produce visually similar results. For decreased \ac{SNR}, the \acs{DL}-based methods achieve superior denoising performance but may introduce artifacts for extremely noisy components. However, since the absolute signal energy in these components is close to zero, their influence on the final reconstruction is negligible. In these cases, the \acs{DCT-F} method tends to retain the noise present in the input data. In the reconstructed images, all three learning-based models provide comparable noise suppression, with only minor differences visible in the upper region of the snake phantom. Overall, reconstructions obtained from denoised \acp{SM} appear slightly more blurred than the ground-truth reference reconstructed from high-\ac{SNR} data.

\subsection{Accelerated calibration}

The quantitative comparison of SMRnet and tricubic interpolation for the accelerated calibration problem is presented in \autoref{table:upsampling_simulated_3d}. The methods show comparable performance, with overall \ac{PSNR} and \ac{SSIM} values remaining relatively low (\ac{SSIM} barely exceeds 0.6, while \ac{PSNR} stays below 22\,dB). Notably, the performance of both methods degrades for \acp{SM} with lower average sizes. SMRnet, which was also trained to suppress noise, shows higher robustness to it.

\begin{table}[]
    \caption{The results for accelerated calibration on simulated 3D data for tricubic interpolation and SMRnet methods. Mean $\pm$ 95\% \ac{CI} of \ac{PSNR} and \ac{SSIM} are reported across different average size ranges and noise levels ($\sigma$).}
    \centering
    \begin{tabular}{lccccc}
\toprule
\multirow{2}{*}{$\sigma$} & \multirow{2}{*}{Avg. size range} & \multicolumn{2}{c}{Tricubic} & \multicolumn{2}{c}{SMRnet} \\
& & PSNR & SSIM & PSNR & SSIM \\ \midrule
\multirow{4}{*}{$0.0$} & (12, 18] & $17.1\pm0.1$ & $0.353\pm0.009$ & $16.6\pm0.1$ & $0.24\pm0.01$ \\
 & (18, 24] & $19.1\pm0.2$ & $0.44\pm0.01$ & $19.4\pm0.2$ & $0.40\pm0.01$ \\
 & (24, 30] & $21.6\pm0.1$ & $0.601\pm0.010$ & $21.5\pm0.1$ & $0.51\pm0.01$ \\
 & all & $18.75\pm0.09$ & $0.452\pm0.006$ & $18.74\pm0.10$ & $0.383\pm0.007$ \\\midrule
\multirow{4}{*}{$0.1$} & (12, 18] & $16.5\pm0.1$ & $0.326\pm0.008$ & $16.5\pm0.1$ & $0.24\pm0.01$ \\
 & (18, 24] & $18.2\pm0.1$ & $0.38\pm0.01$ & $19.2\pm0.2$ & $0.39\pm0.02$ \\
 & (24, 30] & $19.83\pm0.10$ & $0.461\pm0.010$ & $21.7\pm0.1$ & $0.53\pm0.01$ \\
 & all & $17.86\pm0.07$ & $0.392\pm0.005$ & $18.69\pm0.10$ & $0.378\pm0.007$ \\
\bottomrule
\end{tabular}
    \label{table:upsampling_simulated_3d}
\end{table}

\begin{figure}
 \centering
        \includegraphics[width=1.0\textwidth]{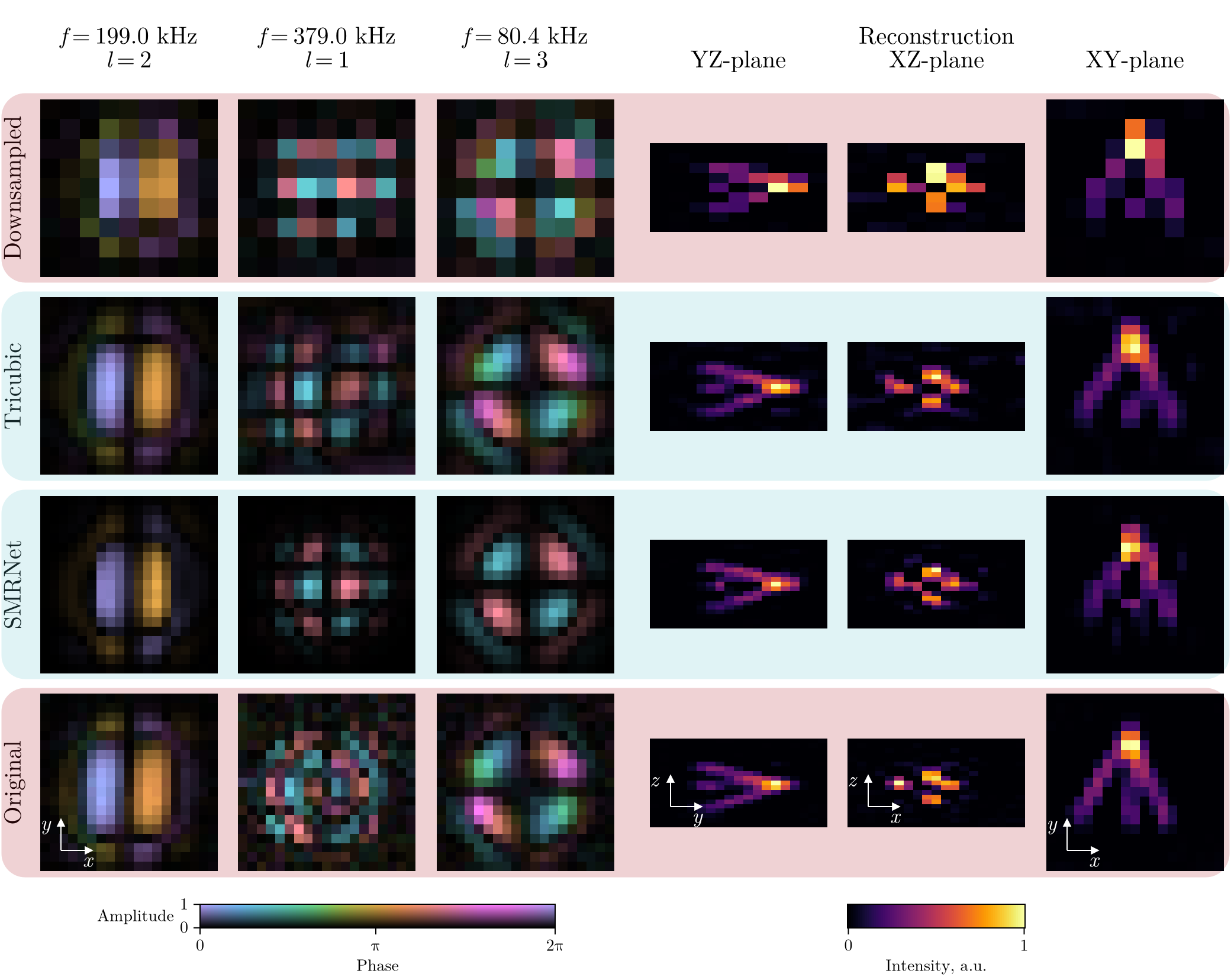}
 \caption{Comparison of tricubic interpolation and SMRnet for restoring a downsampled measured 3D \ac{SM} corresponding to the resolution phantom. Selected frequency components from receive coils $l \in \{1, 2, 3\}$ in $xy$-plane and central slices of the reconstructions in the $xy$-, $xz$-, and $yz$-planes are shown.}
\label{fig:sm-upsampling-real}
\end{figure}

Both methods were applied to the downsampled \ac{SM} of the resolution phantom from the Open~MPI dataset. \autoref{fig:sm-upsampling-real} shows selected frequency components and central slices of the corresponding reconstructions along each spatial axis. Compared with tricubic interpolation, SMRnet reduces noise and alters the phase in the frequency components. The reconstructions are of comparable quality.

\subsection{Upsampling}

\autoref{table:upsampling_simulated_2d} presents the evaluation results on the simulated test set, both with and without added noise, comparing bicubic interpolation and the SMRnet model. For both scaling factors ($\times 2$ and $\times 4$) in the absence of noise, SMRnet consistently outperforms bicubic interpolation by a large margin (approximately \num{25} dB higher \ac{PSNR} and \num{0.08} higher \ac{SSIM}). Although overall performance decreases in the presence of noise, the relative ranking between the methods remains unchanged.

\begin{table}
    \centering
    \caption{Upsampling results on simulated 2D data for bicubic interpolation and SMRnet methods. Mean $\pm$ 95\% \ac{CI} of \ac{PSNR}/\ac{SSIM} are reported across scaling factors and noise levels ($\sigma$).}
    \begin{tabular}{llcccc}
\toprule
\multirow{2}{*}{$\sigma$} & \multirow{2}{*}{Scale} & \multicolumn{2}{c}{Bicubic} & \multicolumn{2}{c}{SMRnet} \\
& & PSNR & SSIM & PSNR & SSIM \\ \midrule
\multirow{2}{*}{$0.0$} & $\times 2$ & $28.4\pm0.4$ & $0.911\pm0.010$ & $54.0\pm0.6$ & $0.993\pm0.002$ \\
 & $\times 4$ & $26.2\pm0.4$ & $0.87\pm0.01$ & $52.4\pm0.7$ & $0.992\pm0.003$ \\\midrule
\multirow{2}{*}{$0.1$} & $\times 2$ & $21.9\pm0.2$ & $0.58\pm0.01$ & $38.9\pm0.3$ & $0.976\pm0.004$ \\
 & $\times 4$ & $21.1\pm0.2$ & $0.47\pm0.01$ & $38.1\pm0.4$ & $0.963\pm0.005$ \\
\bottomrule
\end{tabular}
    \label{table:upsampling_simulated_2d}
\end{table}

\begin{figure}
 \centering
        \includegraphics[width=1.0\textwidth]{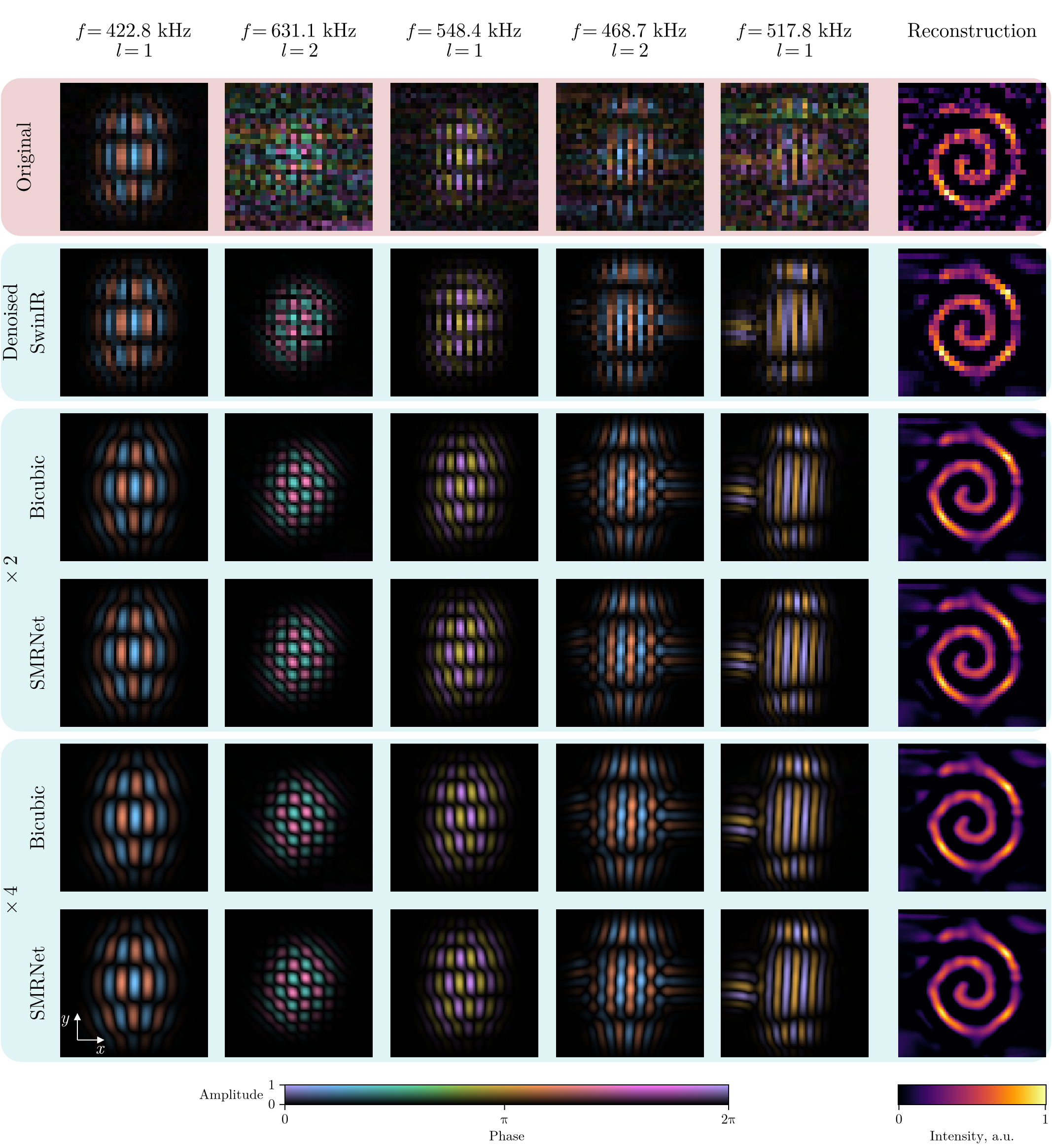}
 \caption{Comparison of bicubic interpolation and SMRnet for upsampling a measured 2D \ac{SM} of the spiral phantom after SwinIR denoising. Selected frequency components from receive coils $l \in \{1, 2, 3\}$ and reconstructions are shown for both methods at factors \num{2} and 4.}
\label{fig:sm-upsampling-real-spiral}
\end{figure}

We evaluate the trained models on measurements of the spiral phantom. The original \ac{SM} with a spatial size of $29\times29$ is first denoised using the SwinIR method and subsequently upsampled to higher resolutions using both bicubic interpolation and the SMRnet model. Selected frequency components and the corresponding reconstructions are shown in \autoref{fig:sm-upsampling-real-spiral}. A noticeable denoising effect is observed in both the frequency components and the reconstructed images. SMRnet demonstrates good generalization to real data, resulting in improved perceived image quality. However, SMRnet and bicubic interpolation exhibit only minor visual differences in both the frequency components and the reconstructions compared to the overall upsampling effect.

\subsection{Inpainting}

\autoref{table:inpainting_simulated} summarizes the quantitative evaluation of inpainting performance on simulated \ac{SM} data with a fixed 10\% mask ratio across average \ac{SM} size ranges and noise levels ($\sigma$). For noiseless \acp{SM} ($\sigma = 0.0$), biharmonic inpainting achieves considerably higher \ac{PSNR} and \ac{SSIM} values (up to \num{34.1} dB and 0.94, respectively). However, its performance degrades substantially with increasing noise (PSNR $\approx$ \num{21.6} dB at $\sigma = 0.1$). In contrast, \acs{PConvUNet}, trained jointly for inpainting and denoising, exhibits greater robustness, maintaining relatively stable performance across noise levels. Furthermore, \acs{PConvUNet} demonstrates improved results for larger \acp{SM}, suggesting a sensitivity to the spatial scale.

\begin{table}[]
    \caption{Inpainting results on simulated \ac{SM} data with 10\% mask ratio. Mean $\pm$ 95\% \ac{CI} of \ac{PSNR}/\ac{SSIM} are reported for biharmonic and PConvUNet methods across \ac{SM} size ranges and noise levels ($\sigma$).}
    \centering
    \begin{tabular}{lccccc}
\toprule
\multirow{2}{*}{$\sigma$} & \multirow{2}{*}{Avg. size range} & \multicolumn{2}{c}{Biharmonic} & \multicolumn{2}{c}{PConvUNet} \\
& & PSNR & SSIM & PSNR & SSIM \\ \midrule
\multirow{4}{*}{$0.0$} & (12, 18] & $30.3\pm0.5$ & $0.930\pm0.004$ & $18.3\pm0.2$ & $0.45\pm0.01$ \\
 & (18, 24] & $31.4\pm0.5$ & $0.936\pm0.004$ & $22.1\pm0.2$ & $0.63\pm0.02$ \\
 & (24, 30] & $37.7\pm0.7$ & $0.959\pm0.003$ & $27.7\pm0.3$ & $0.81\pm0.01$ \\
 & all & $34.1\pm0.3$ & $0.943\pm0.001$ & $22.7\pm0.1$ & $0.645\pm0.007$ \\\midrule
\multirow{4}{*}{$0.1$} & (12, 18] & $21.40\pm0.09$ & $0.847\pm0.005$ & $18.3\pm0.2$ & $0.44\pm0.01$ \\
 & (18, 24] & $21.9\pm0.1$ & $0.795\pm0.008$ & $22.0\pm0.2$ & $0.62\pm0.02$ \\
 & (24, 30] & $22.1\pm0.1$ & $0.627\pm0.008$ & $27.4\pm0.3$ & $0.80\pm0.01$ \\
 & all & $21.60\pm0.04$ & $0.653\pm0.005$ & $22.7\pm0.1$ & $0.641\pm0.006$ \\
\bottomrule
\end{tabular}
    \label{table:inpainting_simulated}
\end{table}

\begin{figure}
 \centering
        \includegraphics[width=1.0\textwidth]{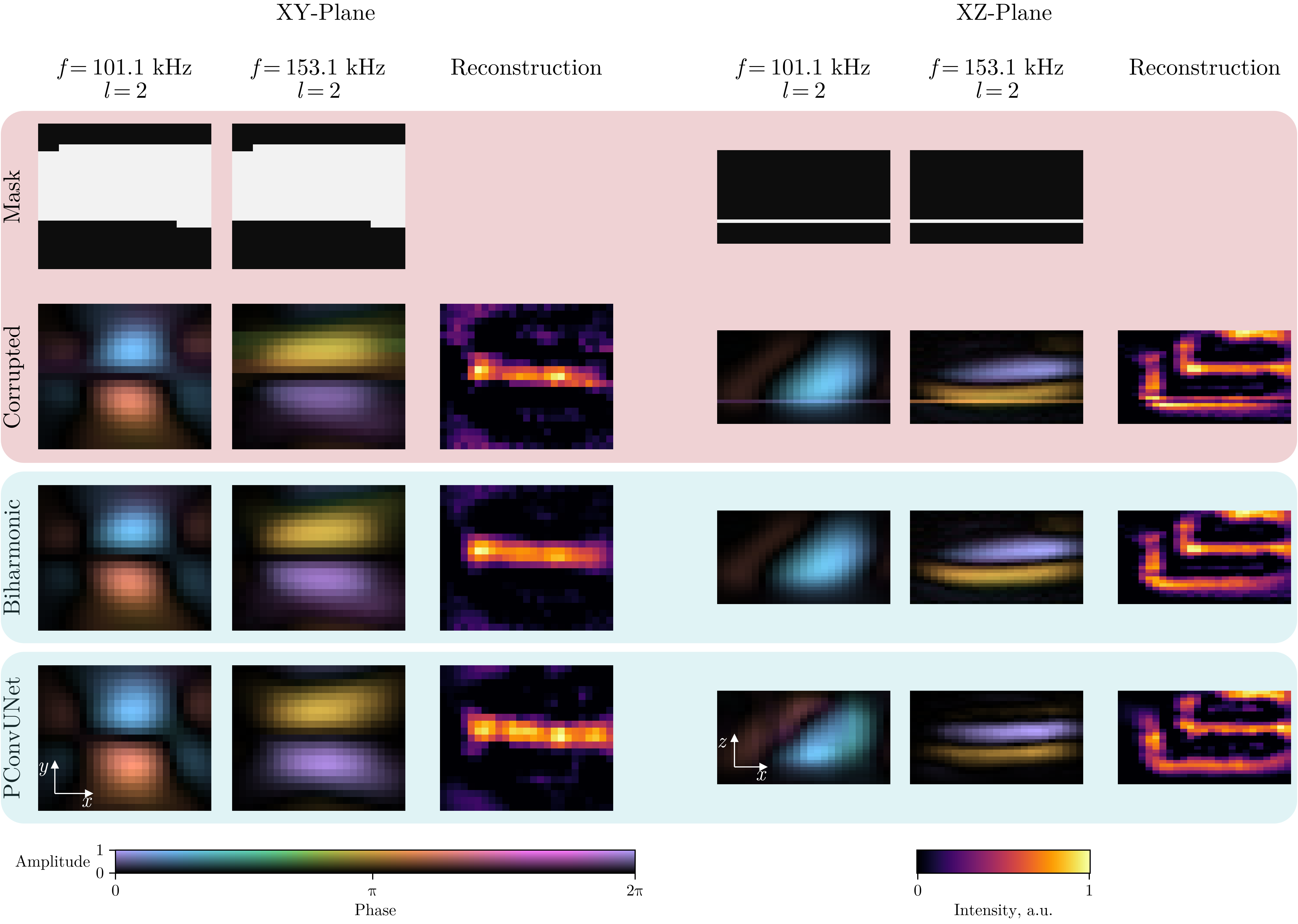}
 \caption{Comparison of biharmonic interpolation and \acs{PConvUNet} for inpainting a corrupted 3D \ac{SM}. Selected frequency components from receive coil $l=2$ and reconstructions are shown in the $xy$- and $xz$-planes for $z=7$ and $y=12$ respectively.}
\label{fig:sm-inpainting-real}
\end{figure}

In addition to the simulated evaluation, both methods were applied to real calibration data. \autoref{fig:sm-inpainting-real} shows two representative frequency components and the corresponding reconstructions for selected planes of the rectangular phantom. The corrupted calibration data exhibit planar artifacts that result in missing pixels in the reconstructed volume. After masking the invalid pixels and performing inpainting, both methods successfully restore the missing information and improve the reconstruction quality. Trained jointly for denoising, the \acs{PConvUNet} also modifies pixels outside the masked regions, which in some cases introduces minor artifacts. Nevertheless, the \acs{PConvUNet} reconstruction appears less blurry than that of the biharmonic interpolation, likely due to its additional denoising capability.

\subsection{Ablation study: impact of simulation model on transferability to real data}

\autoref{table:ablation} presents the results of the ablation study assessing the impact of the simulation model. 
The RDN-Langevin model trained on the dataset generated with the Langevin model does not generalize to the test set containing cases with anisotropy, 
resulting in lower performance compared to the RDN model trained on the dataset generated with the equilibrium model with anisotropy.
The gap in \ac{SSIM} values is more pronounced for higher noise levels, where more prior knowledge about the data distribution is required for effective denoising. 

When applied to the real \ac{SM} of the snake phantom (see~\autoref{fig:sm-denoising-real}), the RDN-Langevin model shows similar behavior:
the RDN-Langevin model tends to produce slightly more symmetric frequency components with the absence of anisotropies, which are present in the RDN model and the original \ac{SM}. 
The resulting reconstruction using \ac{SM} obtained with the RDN-Langevin model appears more noisy. 

\begin{table}[]
    \caption{
        Comparison of 2D system-matrix denoising performance for the RDN and RDN-Langevin models. 
        The models are trained on simulated data generated with the equilibrium model with anisotropy and the Langevin equilibrium model, respectively. 
        Reported values are mean $\pm$ 95\% \ac{CI} of \ac{PSNR}/\ac{SSIM} over noise levels ($\sigma$) on the test set described in \autoref{sec:data_generation}, excluding immobilized cases.
    }
    \centering
    \begin{tabular}{lccr@{ $\pm$ }lr@{ $\pm$ }l}
\toprule
 & $\sigma$ & Metric & \multicolumn{2}{c}{RDN} & \multicolumn{2}{c}{RDN-Langevin} \\
\midrule
\multirow{2}{*}{} & \multirow{2}{*}{0.06} & PSNR & $39.6$ & $0.4$ & $31.3$ & $0.5$ \\
 &  & SSIM & $0.992$ & $0.001$ & $0.94$ & $0.01$ \\
\addlinespace[0.3em]
\multirow{2}{*}{} & \multirow{2}{*}{0.10} & PSNR & $36.8$ & $0.4$ & $29.3$ & $0.5$ \\
 &  & SSIM & $0.983$ & $0.003$ & $0.91$ & $0.01$ \\
\addlinespace[0.3em]
\multirow{2}{*}{} & \multirow{2}{*}{0.20} & PSNR & $32.2$ & $0.4$ & $26.2$ & $0.4$ \\
 &  & SSIM & $0.959$ & $0.006$ & $0.86$ & $0.02$ \\
\addlinespace[0.3em]
\multirow{2}{*}{} & \multirow{2}{*}{0.30} & PSNR & $29.5$ & $0.4$ & $24.4$ & $0.4$ \\
 &  & SSIM & $0.929$ & $0.008$ & $0.81$ & $0.02$ \\
\addlinespace[0.3em]
\bottomrule
\end{tabular}
    \label{table:ablation}
\end{table}

\section{Discussion}

This study shows that simulated \acp{SM} are suitable for training \ac{DL} models for multiple \ac{SM} restoration tasks and that these models qualitatively generalize to measured data. The largest and most robust gains over non-learning baseline methods were observed for denoising, where learning-based methods improved both frequency components and reconstructions. Upsampling yielded modest improvements on measured data, whereas accelerated calibration and inpainting benefited most when combined with denoising, which also provided phase correction and noise suppression.

Quantitative evaluation on measured \acp{SM} is limited by the lack of \ac{GT} and the small number of available datasets. 
Consequently, quantitative analyses were performed on simulations, acknowledging a potential domain shift between simulated and measured distributions. 
On simulations, learning-based methods clearly outperformed classical baseline methods for denoising and upsampling; 
however, for upsampling the numerical advantage did not consistently translate into perceptual gains on measurements. 
Two factors likely contribute: 
(i) de‑aliasing patterns learned by the model are partially obscured by measurement noise, and 
(ii) a preceding denoising step can shift the data distribution. 
For accelerated calibration, performance was strongly dependent on the calibration grid size and noise level. 
Tricubic interpolation showed minor improvements over SMRnet in the absence of noise, but this trend reversed for noisy \acp{SM}, where SMRnet proved more robust. The performance gap between the methods under noisy conditions increased with calibration grid size.
Since measured \acp{SM} always contain noise and accelerated calibration is particularly relevant for larger calibration grids, 
the \ac{DL}-based method may offer greater practical value.
Despite low \ac{PSNR} and \ac{SSIM} values, the reconstructions of the resolution phantom in \autoref{fig:sm-upsampling-real} sufficiently captured the phantom structure. 
This suggests that pixel-wise performance metrics measured for frequency components do not fully reflect the quality of the reconstructed images, 
as the reconstruction process itself can compensate for imperfections in the \ac{SM}.
For inpainting, a similar behavior was observed: the learning-based method outperformed the classical approach for noisy \acp{SM} and larger calibration grids.
The results indicate that classical methods may still be effective for frequency components with high \ac{SNR}.
However, for the entire \ac{SM}, which typically exhibits varying \ac{SNR} across frequencies, the learning-based method can provide more consistent performance across frequencies as 
it is more robust to noise.
For both accelerated calibration and inpainting, future works may explore size-aware conditioning or stratified training to further improve 
performance across varying calibration grid sizes.

Reconstruction can compensate for imperfections in the \ac{SM} through priors (e.g., Tikhonov regularization) and \ac{SNR}-based frequency selection, 
which can mask the effects of SM restoration. 
For this reason, the reconstruction parameters were fixed across methods for each measurement. 
Otherwise, any improvement cannot be attributed specifically to the \ac{SM} restoration method, because the reconstruction also affects the result. 
Nonetheless, restored \acp{SM} enable lowering the \ac{SNR} threshold and regularization strength while maintaining image quality, 
thereby reducing bias and leveraging more frequency components. 
Additionally, for measurements with low \ac{SNR}, recovering the \ac{SM} components at frequencies with low signal may not improve reconstruction quality.
However, if the measurements have high \ac{SNR}, the method gives the ability to increase \ac{SNR} of a \ac{SM} without performing additional averaging or sampling more calibration points, thus reducing the \ac{SM} calibration time.

As \ac{MPI} remains a preclinical imaging modality, available measurements are limited and do not permit a rigorous quantitative evaluation of restoration-induced improvements in image quality. Nevertheless, all cases considered in this work offer potential reductions in the calibration time when restoration methods are applied. Hereafter, the calibration time is estimated by neglecting robot motion and computing the product of the number of grid positions, the repetition time, and the number of averages. In this study, the repetition times are \qty{652.8}{\micro s} for the 2D sequence and \qty{21.54}{ms} for the 3D sequence.
For the snake phantom (denoising), the calibration time could be reduced from \qty{55}{s} to \qty{2}{s}, because it scales linearly with the number of averages.  For the resolution phantom (accelerated calibration), the calibration time could be reduced from \qty{4.1}{h} to \qty{31}{min}. For the rectangular phantom (inpainting), recovery of the corrupted \ac{SM} would save \qty{4.2}{h}, since such measurements would otherwise typically be discarded.

In this study, the equilibrium model with uniaxial anisotropy was selected as it offered a favorable trade-off between computational cost and applicability to real data. Other simulation methods, such as Langevin or Fokker-Planck, can also be used for data generation \parencite{ALBERS2022168508, knopp_exploiting_2023, kayapinar_fourier_2024, faldum_efficient_2025}. However, under extreme corruption levels, models may behave generatively and introduce artifacts. If relevant physical effects are not included in the simulation, generalization to measurements exhibiting those effects is impaired, as shown by the ablation study.

The method was restricted to 2D and 3D Lissajous trajectories on an \ac{FFP} scanner with specific \ac{DF} frequencies, but the approach can be extended to other sequences or systems, such as Cartesian trajectories or \ac{FFL} systems. While background noise is scanner-specific, it can be re-measured via empty-frame acquisitions, making the procedure practical across systems; the simulated \ac{GT} \acp{SM} remain transferable. Although measured noise characterizes the system most accurately, simplified parametric noise models may be sufficient for model generalization and remain a topic for future work.

Data-efficient models trained on small amounts of real data may outperform simulation-trained models in narrow domains. The proposed approach is most valuable when self-supervised learning is infeasible or \ac{GT} is unavailable (e.g., denoising and upsampling). Additionally, large simulated datasets enable pre-training of high-capacity models that can later be fine-tuned on limited real data.

Future work may address the need for a comprehensive collection of \ac{MPI} \ac{SM} measurements to enable rigorous quantitative evaluation of data-driven methods, as well as extend the proposed approach to additional acquisition trajectories and scanner types.

\section{Conclusion}

We have found that simulated \acp{SM} can be a valid source of training data for \ac{DL} methods solving \ac{SM} restoration problems. The proposed data generation method has the potential to improve the quality of existing \ac{DL} methods that suffer from data scarcity or develop new methods for the problems where \ac{GT} data were not available before.

\data{The data that support the findings of this study are openly available at \url{https://doi.org/10.15480/882.16265} \parencite{tsanda_data_supplement}. }

\ack{
We thank Martin~Möddel for his valuable comments and feedback during the preparation of this manuscript.
}

\funding{This project is funded by the Deutsche Forschungsgemeinschaft
(DFG, German Research Foundation) - SFB 1615 - 503850735.}




\printbibliography

\end{document}